\documentclass[twocolumn,aps,preprintnumbers,prb,showpacs%
,amsfonts,amsmath,amssymb,superscriptaddress,floatfix,dblfloatfix%
,10pt,lettersize]{revtex4}

\usepackage[dvips]{graphicx}
\usepackage{bm}

\graphicspath{{eps/}}

\usepackage{color}

\newcommand{\affA}{%
Department of Applied Physics, School of Engineering,
The University of Tokyo,\\
7-3-1 Hongo, Bunkyo-ku, Tokyo, 113-8656, Japan}

\newcommand{\affB}{%
Quantum ICT Research Institute, Tamagawa University,\\
6-1-1 Tamagawa Gakuen, Machida, Tokyo, 194-8610, Japan}

\newcommand{\affC}{%
School of Physics and Astronomy, University of Southampton,\\
Southampton SO17 1BJ, United Kingdom}

\newcommand{\affD}{%
NTT Device Technology Laboratories,\\
3-1, Morinosato, Wakamiya, Atsugi, Kanagawa, 243-0198, Japan}

\newcommand{\affE}{%
Centre for Quantum Photonics, H. H. Wills Physics Laboratory and Department of Electrical and Electronic Engineering, University of Bristol,\\
Merchant Venturers Building, Woodland Road, Bristol, BS8 1UB, United Kingdom}

\begin{document}

\title{
Continuous variable entanglement on a chip}

\date{\today}

\author{Genta Masada}
\affiliation{\affA}
\affiliation{\affB}
\author{Kazunori Miyata}
\affiliation{\affA}
\author{Alberto Politi}
\affiliation{\affC}
\author{Toshikazu Hashimoto}
\affiliation{\affD}
\author{Jeremy L. O'Brien}
\affiliation{\affE}
\author{Akira Furusawa}
\affiliation{\affA}

\pacs{03.67.Hk, 42.50.Dv, 42.65.Yj}

\maketitle

{\bf
\noindent
Encoding  quantum information in continuous variables (CV)---as the quadrature of electromagnetic fields---is a powerful approach to quantum information science and technology\cite{OBrien09}. CV entanglement---light beams in Einstein-Podolsky-Rosen (EPR)\cite{Einstein35} states---is a key resource for quantum information protocols\cite{Braunstein05}; and enables hybridisation between CV and single photon discrete variable (DV) qubit systems\cite{Furusawa11}. However, CV systems are currently limited by their implementation in free-space optical networks: increased complexity, low loss, high-precision alignment and stability, as well as hybridisation, demand an alternative approach. Here we show an integrated photonic implementation of the key capabilities for CV quantum technologies---generation and characterisation of EPR beams in a photonic chip. Combined with integrated squeezing and non-Gaussian operation, these results open the way to universal quantum information processing with light.
}

The use of quantum mechanical systems to encode, transmit and manipulate information promises the development of disruptive technologies in the fields of sensing, computing and communications\cite{Dowling03}.  There are two natural ways to encode information in the quantum domain: using two (or more) discrete levels of a quantum system---\emph{e.g.} the spin of an electron or the polarization of a photon; or using a continuum of states---\emph{e.g} a quantized harmonic oscillator---which can be described by continuous variables, such as the position and momentum of a particle. Light is particularly suited to encode information in the quantum domain and has been successfully used to demonstrate quantum protocols both with discrete variables (DV) using the degrees of freedom of single photons  \cite{Bouwmeester97, OBrien07} and continuous variables (CV) using the quadratures of beams of light \cite{Furusawa98, Weedbrook12}. 

DV systems benefit from high fidelity operations, but are currently limited by imperfect generation and detection of single photons, and the absence of deterministic interaction of photons. In contrast, CV schemes can achieve deterministic, unconditional operation but admit lower fidelities for the majority of quantum protocols. A new hybrid approach overcomes these problems 
by combining the benefits of both DV and CV in a single system \cite{Furusawa11}. Already entanglement between DV and CV systems  \cite{Jeong14, Morin14} and teleportation of a DV state using CV protocols \cite{Takeda13} has been demonstrated.
However, practical application of this hybrid approach will require an integrated system where light is guided, manipulated and made to interfere in a waveguide architecture. Integrated waveguide circuits have been successfully developed for DV applications \cite{Politi08,Mathews09} demonstrating high fidelity and reliable operation. A benefit of this optical integration, in addition to miniaturization of quantum circuits, is attainment of a high degree of spatial mode matching---essential for classical and quantum interference---without any optical adjustment \cite{Shaodbolt12}. Transferring CV systems to such integrated circuits presents considerable technical challenges, including high-efficiency coupling, low loss operation, and a high degree of mode matching, in an architecture compatible with DV operation.

Here we report the demonstration of an integrated circuit that can be reconfigured to implement the fundamental operations required for CV quantum optics, compatible with  high fidelity DV operation\cite{Laing10}.
We characterize a device composed of directional couplers that can interfere two squeezed states of light
, perform phase stabilization and mix strong local oscillator (LO) beams to perform homodyne detection. 
We further configure the integrated optical circuit to generate and characterize Einstein-Podolsky-Rosen (EPR) entangled beams of light, obtaining a correlation variance $\Delta ^{2} _{\rm 1,2}=0.71$ that demonstrates inseparability. 
This shows all the components required to achieve quantum teleportation \cite{Furusawa98}, which is a Gaussian operation and the most fundamental quantum
information protocol, on a single integrated chip.

\begin{figure*}
\centerline{\includegraphics[width=14cm,clip]{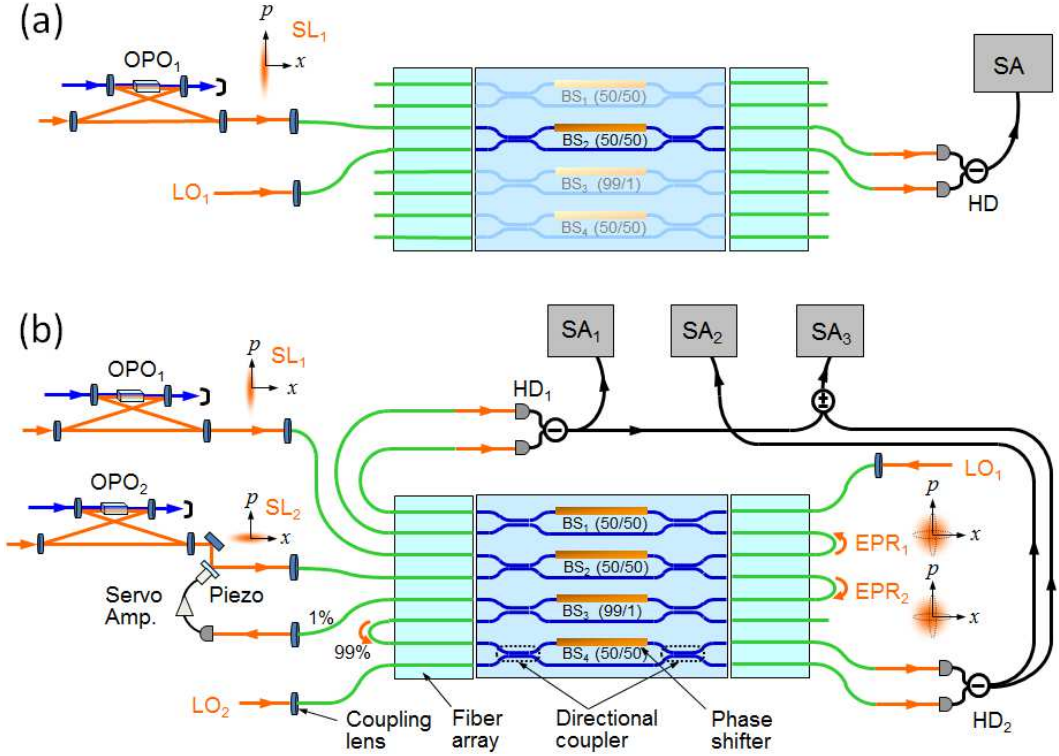}}
\caption{Experimental setup of (a)balanced homodyne measurement of squeezed light and (b)generation and characterization of EPR beams within a photonic chip.
The beam splitter BS$_{\rm{1}}$, BS$_{\rm{2}}$, and BS$_{\rm{4}}$ are tuned as 50 to 50 and the BS$_{\rm{3}}$ is tuned as 99 to 1 of splitting ratio.
A pair of fiber arrays are used for both entrance and exit facets of the chip to inject and eject beams efficiently.
In order to guide the output from one beam splitter BS to another one, two fibers are mechanically connected.
Continuous wave squeezed lights SL$_{\rm{1}}$ and SL$_{\rm{2}}$ at 860 nm are generated by sub-threshold optical parametric oscillators (OPOs)\cite{Suzuki06,Takeno07} outside the chip.
A pump beam at 430 nm for OPOs is generated by an optical frequency doubler\cite{Masada10}.
In  the setup (a) the squeezed light SL$_{\rm{1}}$ is combined with the local oscillator LO$_{\rm{1}}$ and detected by balanced homodyen detector HD outside the chip.
In the setup (b), firstly two squeezed lights SL$_{\rm{1}}$ and SL$_{\rm{2}}$ are combined at beam splitter BS$_{\rm{2}}$.
In our experiment weak coherent beams are introduced into both OPOs.
1\% of the weak coherent beams is picked up by beam splitter BS$_{\rm{3}}$ and used for phase locking between squeezed lights SL$_{\rm{1}}$ and SL$_{\rm{2}}$ at 90 degrees by using servo-amplifier and piezo actuator.
The output beam EPR$_{\rm{1}}$(EPR$_{\rm{2}}$) is combined with local oscillator LO$_{\rm{1}}$(LO$_{\rm{2}}$) at beam splitter BS$_{\rm{1}}$(BS$_{\rm{4}}$), and then detected by balanced homodyne detector HD$_{\rm{1}}$(HD$_{\rm{2}}$). 
}

\label{setup}
\end{figure*}

We first demonstrate that photonic circuits can be used in CV quantum optics and provide the required suppression of coupling between different waveguide modes. Unwanted cross-coupling is typically given by stray-light that arises from insufficient coupling from fiber inputs, waveguides bend losses and cladding guiding. Such cross-coupling is a major limitation  for quantum optical networks, reducing the fidelities of quantum operations. It is particularly troublesome for balanced homodyne measurements, since quantum information is encoded on weak optical signals, and the presence of light scattered from intense coherent beams such as the LO can degrade the signal and destroy non-classical effects. Here we avoid the harmful consequences of stray light by using the side-bands of the optical field and  almost perfect mode matching between squeezed light (SL) and LO within a chip, as shown below.

Figure \ref{setup} shows the experimental setup, where the optical network consists of a silica on silicon chip that includes four waveguide interferometers (See Supplementary Information for details). The interferometers are composed of a pair of directional couplers and a phase shifter implemented by a resistive heater lithographically defined on top of one waveguide arm \cite{Mathews09}.
By tuning the drive current of each of the resistive heaters we control the optical phase difference between the two arms. Each interferometer is thereby equivalent to a tunable beam splitter (BS).
As shown in Fig.~\ref{setup}(a), the squeezed light SL$_{\rm{1}}$ and local oscillator LO$_{\rm{1}}$ are coupled into the chip and superimposed at beam splitter BS$_{\rm{2}}$.
Output beams are collected by optical fibers and measured by a balanced homodyne detector outside the chip. We performed the experiment at a side-band of 1.5 MHz from the laser frequency. 

\begin{figure}[h]
\centerline{\includegraphics[width=7.5cm,clip]{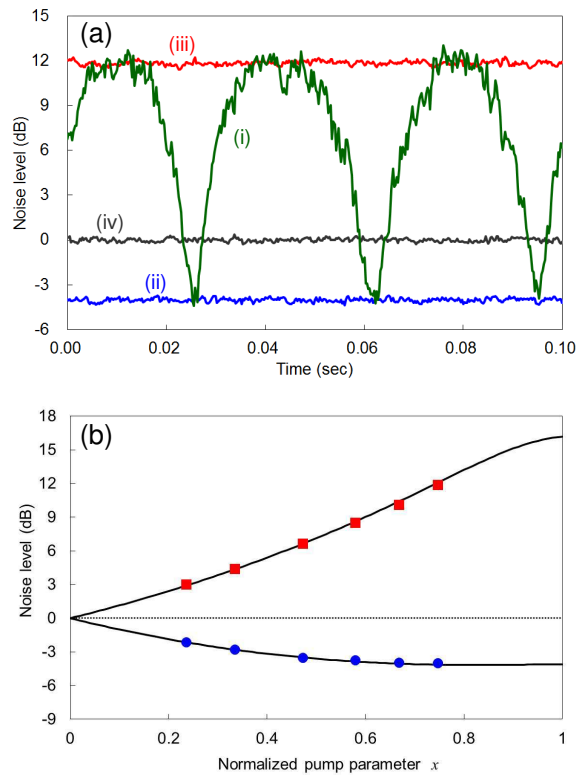}}
\caption{On-chip homodyne detection of squeezing. (a) Quantum noise level of squeezed light generated with a pump power of 100 mW.
(i)LO phase is scanned, (i\hspace{-.1em}i)LO phase is locked at the squeezed quadrature, (i\hspace{-.1em}i\hspace{-.1em}i)LO phase is locked at the antisqueezed quadrature, and (i\hspace{-.1em}v)shot noise level normalized to 0 dB.
Resolution bandwidth is 30 kHz and video bandwidth is 300 Hz.
Traces (i\hspace{-.1em}i), (i\hspace{-.1em}i\hspace{-.1em}i) and (i\hspace{-.1em}v) are averaged 20 times.
(b) Pump power dependence of the squeezing and antisqueezing levels.
The normalized pump power $x$ is defined as $\sqrt{P/P_{th}}$ where 
$P_{th}=179$ mW is the oscillation threshold of the OPO.
Circles/squares indicate observed squeezing/antisqueezing levels, solid curves represent the result of numerical calculations.
}
\label{SQresult}
\end{figure}

Figure \ref{SQresult}(a) shows typical results of homodyne measurement of the squeezed light SL$_{\rm{1}}$, which is generated by using a sub-threshold optical parametric oscillator (OPO) at a pump power of 100 mW.
As the local oscillator phase is varied using a piezo-electric controller, the observed signal oscillates between a squeezing level of $-4.02 \pm 0.13$ dB and anti-squeezing level of $+11.85 \pm 0.15$ dB.
We repeated the measurement for different pump powers. Figure~\ref{SQresult}(b) shows the dependence of squeezing and antisqueezing levels as the pump power is varied. The solid curves represent calculation results (See Supplementary Information) with almost perfect visibility (=0.995) and show good agreement with the experimental values. The saturation of the squeezing level around $-4$ dB at higher pump power is predominantly due to insufficient overall coupling efficiency (=0.72), consistent with the $-8.4$ dB of squeezing level expected for perfect coupling efficiency. These results, and their agreement with calculations, demonstrate high performances of our integrated device for CV operation. In particular, a high degree of  mode matching (interference between two coherent beams shows 0.995 visibilities) and low levels of stray light noise (owing to the side-band measurement) enable the transmission and detection of high levels of squeezing.

We now describe the central result of this work: the generation and characterization of  EPR beams within the chip.
In order to characterize the amount of entanglement obtained with our system, we use an inseparability criterion. If we introduce the quadrature phase amplitude operators $\hat{x}$ and $\hat{p}$ that corresponds to cosine and sine component of optical field, 
we can write the complex amplitude operators of each output modes EPR$_{\rm{1}}$ and EPR$_{\rm{2}}$  as $\hat{a}_{\rm{1}} = \hat{x}_{\rm{1}} + i\hat{p}_{\rm{1}}$ and $\hat{a}_{\rm{2}} = \hat{x}_{\rm{2}} + i\hat{p}_{\rm{2}}$, respectively.
To evaluate the entanglement between two output beams, we define the correlation variance $\Delta ^{2} _{\rm 1,2}$ as
\begin{equation}
\Delta ^{2} _{\rm 1,2} = \langle [\Delta (\hat{x}_{\mathrm{1}}-\hat{x}_{\mathrm{2}})]^2\rangle +\langle [\Delta (\hat{p}_{\mathrm{1}}+\hat{p}_{\mathrm{2}})]^2\rangle.
\label{criterion}
\end{equation}
It is proven by Duan \textit{et al}.\cite{Duan00} and Simon\cite{Simon00} that two beams are inseparable when $\Delta ^{2} _{\rm 1,2} < 1$, demonstrating entanglement. 
It is also possible to re-write $\Delta ^{2} _{\rm 1,2}$ as a function of the squeezing parameter $r$ of the input beams, in this case we have $\Delta ^{2} _{\rm 1,2} = e^{-2r}$. It can be easily seen that $\Delta ^{2} _{\rm 1,2}$ goes to zero in the ideal case with infinite amount of squeezing, achieving perfect correlation.

\begin{figure*}
\centerline{\includegraphics[width=14cm,clip]{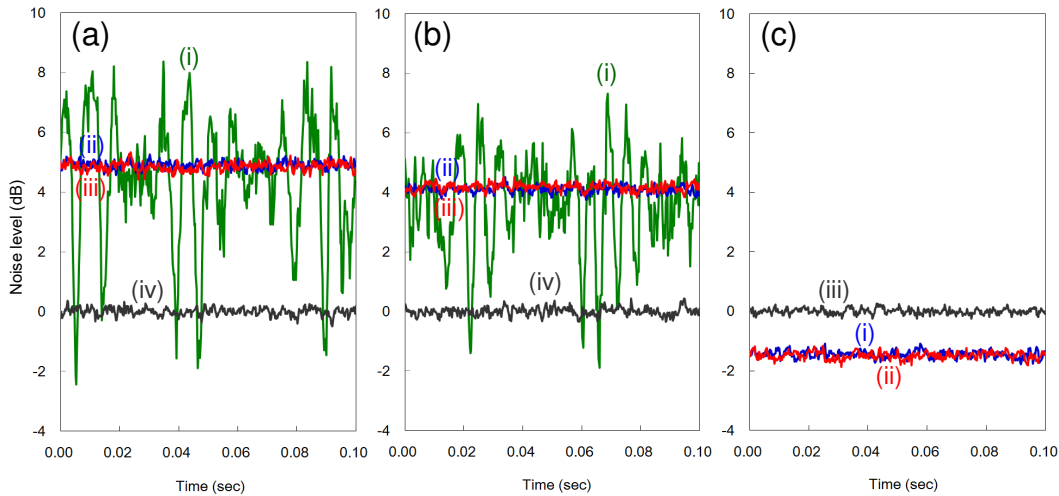}}
\caption{
EPR entanglement verification. Noise levels for (a) output beam EPR$_{\mathrm{1}}$ and (b) output beams EPR$_{\mathrm{2}}$.
(i)noise measurement changing $\theta_{\rm{1,2}}$ and simultaneously scanning the LO phase with a faster timescale, (i\hspace{-.1em}i),(i\hspace{-.1em}i\hspace{-.1em}i)LO phase is locked at $x$ and $p$ quadrature with the $\theta_{\rm{1,2}}$ locked at 90 degrees, and (i\hspace{-.1em}v)shot noise level normalized to 0 dB.
(c) Inseparability condition measurement
(i) and (i\hspace{-.1em}i) represent $\langle [\Delta (\hat{x}_{\mathrm{1}}-\hat{x}_{\mathrm{2}})]^2\rangle$ and $\langle [\Delta (\hat{p}_{\mathrm{1}}+\hat{p}_{\mathrm{2}})]^2\rangle$ respectively, and (i\hspace{-.1em}i\hspace{-.1em}i) is the noise level without quantum correlation.
}
\label{EPRresult}
\end{figure*}

Experimentally we generate the EPR beams combining two squeezed lights SL$_{\rm{1}}$ and SL$_{\rm{2}}$ at beam splitter BS$_{\rm{2}}$ and locking the relative phase between them $\theta_{\rm{1,2}}$ at 90 degrees (see Fig.\ref{setup}(b))\cite{Furusawa11}.
Quantum entanglement is verified with homodyne measurement, using beam splitters BS$_{\rm{1}}$ and BS$_{\rm{4}}$ to combine the two EPR beams with two LO beams. Also in this case the experiment is performed at 1.5 MHz side-band of the laser frequency.
Fig.\ref{EPRresult}(a) and (b) show the noise levels of the output beams EPR$_{\mathrm{1}}$ and EPR$_{\mathrm{2}}$, respectively, as the phase of the LO is varied.
Trace (i) represents the noise of one of the output beams while we vary $\theta_{\rm{1,2}}$ and simultaneously scan the LO phase with a faster timescale.
The outputs continuously tune from EPR beams to two independent squeezed light beams as $\theta_{\rm{1,2}}$ changes from 90 to 0 degrees\cite{Zhang03}.
Traces (i\hspace{-.1em}i) and (i\hspace{-.1em}i\hspace{-.1em}i) represent the noise levels of the $x$ and $p$ quadratures when $\theta_{\rm{1,2}}$ is locked at 90 degrees.
We observed the noise levels of $\langle \Delta {\hat{x}_{\mathrm{1}}}^2\rangle=+4.91 \pm0.13$ dB and $\langle \Delta {\hat{p}_{\mathrm{1}}}^2\rangle=+4.85 \pm0.15$ dB for output EPR$_{\rm{1}}$ at HD$_{\mathrm{1}}$ and of $\langle \Delta {\hat{x}_{\mathrm{2}}}^2\rangle+4.10 \pm0.12$ dB and $\langle \Delta {\hat{p}_{\mathrm{2}}}^2\rangle=+4.18 \pm0.13$ dB for output EPR$_{\rm{2}}$ beam at HD$_{\mathrm{2}}$ above the shot noise level represented by trace (i\hspace{-.1em}v).
This phase insensitive behavior is one of the essential properties of EPR beams: entangled states show strong correlations between each other, but the individual properties are not well defined (tracing out either beam of a maximally entangled state leaves the other beam in a maximally mixed state).

Finally, we measure $\langle [\Delta (\hat{x}_{\mathrm{1}}-\hat{x}_{\mathrm{2}})]^2\rangle$ and $\langle [\Delta (\hat{p}_{\mathrm{1}}+\hat{p}_{\mathrm{2}})]^2\rangle$ to check the inseparability criterion.
Trace (i) in Fig.\ref{EPRresult}(c) shows the variance of difference signal between $\hat{x}_{\rm{1}}$ from HD$_{\rm{1}}$ and $\hat{x}_{\rm{2}}$ from HD$_{\rm{2}}$ which are taken by using a hybrid junction.
Trace (i\hspace{-.1em}i) represents the variance of the sum signal between $\hat{p}_{\rm{1}}$ and $\hat{p}_{\rm{2}}$ obtained in the same way.
We achieve $\langle [\Delta (\hat{x}_{\mathrm{1}}-\hat{x}_{\mathrm{2}})]^2\rangle = -1.44 \pm0.12$ dB and $\langle [\Delta (\hat{p}_{\mathrm{1}}+\hat{p}_{\mathrm{2}})]^2\rangle = -1.49 \pm0.12$ dB below the noise level without quantum correlation shown by trace (i\hspace{-.1em}i\hspace{-.1em}i).
These results yield a correlation variance of $\Delta ^{2}_{\rm 1,2}=0.71$ and satisfy the inseparability criterion, proving the generation of EPR beams and characterization of quantum entanglement within our photonic chip.


Our demonstration of all of the key components for Gaussian operations within a photonic chip, including the generation and characterization of Einstein-Podolsky-Rosen (EPR) beams and the verification of entanglement, points the way to full optical integration of CV and hybrid quantum information processing. The use of side-bands of the optical field---to circumvent noise caused by stray light---combined with high quality interference owing to the perfect mode-matching in directional couplers, enables simultaneous operation on non-classical light and intense coherent beams in a single photonic chip. Coupling in and out of the device is the major contributor to loss, which limits entanglement; this can be dramatically reduced by concatenating circuits in a single integrated network. Further performance gains will be achieved by integration of photodiodes on the photonic chip---as has been achieved with waveguide coupled single photon detectors \cite{Sprengers11,Pernice12}---further removing the requirement of out-coupling from the chip to fibers. Universal quantum information processing will require non-Gaussian operations like photon counting \cite{Dakna97}. Integrated semiconductor  \cite{{Campbell07}} or superconducting \cite{Pernice12} detectors are promising; and will enable side-band techniques using the recently developed optical high pass filter \cite{Takeda12}, thereby eliminating the influence of stray light that would affect the detection dark counts. 

We thank Hans Bachor for helpful advice. This work is partly supported by PDIS, GIA, and APSA commissioned by the MEXT, and NTT.

\vspace{10mm}
\noindent
\textbf{supplementary information}

We provide here supplementary information for our letter including details of the experimental setup and analysis of the squeezed light level.

\section{Experimental setup}
The photonic chip was fabricated on a 4-inch silicon wafer, onto which a 16 $\mu$m layer of thermally grown undoped silica was deposited to form the lower cladding of the waveguides. A 3.5 $\mu$m layer of silica doped with germanium and  boron oxides was then deposited by flame hydrolysis and patterned in 3.5$ \mu$m-wide waveguides using optical lithography. A 16 $\mu$m layer of silica, doped with phosphorous and boron so as to be index-matched with the lower layer, constituted the upper cladding. Simulations indicated single-mode operation at 860 nm. A final metal layer was lithographically patterned on top of the devices to form resistive elements and metal connections and contact pads. The total length of the chip is 26 mm. The chip contains five interferometers consisting of a pair of directional couplers and a phase shifter implemented with a resistive heater. By tuning the drive current in the resistor it is possible to control the phase of one of the optical arms. The waveguide interferometer is acting as a beam splitter (BS) with variable splitting ratio, with reflectivity continuously changeable through the resistive heater.

The waveguides are separated by 250 $\mu$m in the facet region to allow coupling through fiber arrays. Single-mode, polarization-maintaining fibers are arranged on a V-grooved Si substrate with regular intervals to form the fiber array.
Diameters for the core and clad of the fibers are 5 and 125 $\mu$m respectively. 
The gaps between the fiber arrays and chip are filled with the index matching fluid to eliminate reflection losses and increase coupling efficiencies.

We use a continuous-wave Ti:Sapphire laser (Coherent MBR-110) at 860 nm as a light source.
All beams, squeezed lights and local oscillators, are coupled in the fibers through end faces which are designed as FC/APC and antireflection coated at 860 nm.
In order to improve the fiber coupling efficiency, aspheric lenses with focusing length of 8 mm and numerical aperture of 0.5 are used to focus the beam to a waist comparable to the fiber core size.
An overall coupling efficiency $\eta_{\rm{c}}$=0.72 is measured launching a weak coherent beam into the fiber, couping in the photonic chip, and collecting in the exit fiber.
The total efficiency is given as $\eta_{\rm{c}}$=$\eta_{\rm{f}}$*$\eta_{\rm{w}}^{2}$ where $\eta_{\rm{f}}$ is the coupling efficiency in the fiber and $\eta_{\rm{w}}$ is coupling efficiency between fiber and waveguide.
We measure $\eta_{\rm{f}}$=0.87, therefore $\eta_{\rm{w}}$ can be estimated as 0.91 on average if propagation losses are negligible.

\section{Theoretical evaluation for squeezing and antisqueezing level}

We provide supplementary information for the on-chip balanced homodyne measurement of squeezed light which is shown in Figure 2 of the main text.
OPO$_{\rm{1}}$ has a bow-tie configuration with a round trip length of 500 mm,  output coupler transmission $T$=0.113, and contains a periodically poled KTP crystal as a nonlinear optical medium. 
The solid line curves in Figure 2(b) represent the theoretical values of the noise level ${R^{\prime\prime}_\pm}$ for the anti-squeezed ($+$) and squeezed ($-$) quadrature including experimental imperfections as the normalized pump power $x=\sqrt{P/P_{th}}$ is varied. We can write \cite{Takeno07}
\begin{equation}
R_\pm=1\pm\rho\eta \frac{4x}{(1\mp x)^2+4f^2},
\label{Eq:SQ}
\end{equation}
where $f$ is the measurement frequency normalized to the full width at half maximum (FWHM) of the cavity line width, $\rho$ is the escape efficiency and $\eta$ is the homodyne detection efficiency.
In our experiment the FWHM of the cavity can be estimated as 11.8 MHz and the measurement frequency side band mode is 1.5 MHz, so we calculated $f$=0.127.
The homodyne detection efficiency $\eta$ is described as $\eta=\eta_{\rm{PD}}$*$\eta_{\rm{p}}$*$\eta_{\rm{c}}$*$\eta_{\rm{v}}^2$ where $\eta_{\rm{PD}}$ is quantum efficiency of the photodiode, $\eta_{\rm{p}}$ is propagation efficiency of optical pass between the OPO output coupler and fiber coupling lens, and $\eta_{\rm{v}}$ is visibility between the OPO output mode and the local oscillator in the waveguide. 
In our experimental setup, $\eta_{\rm{PD}}$ = 0.998 (Hamamatsu, S3590-06 with antireflection coating), and $\eta_{\rm{p}}$=0.99. 
We estimate $\eta_{\rm{v}}$ by observing the interference signal obtained introducing two weak coherent beams into the chip and obtain $\eta_{\rm{v}}$=0.995.
These results yield a final value of $\eta$=0.704. 
The escape efficiency $\rho$ of the OPO is given as $T$/($T$+$L$) where $L$ is intracavity loss. 
The loss $L$ increases with the pump beam power, as expected by blue light induced infrared absorption (BLIIRA) in the KTP crystal. 
The experimental result can be expressed as $L=L_0+a*P$ where $L_0$ is a passive loss without the pump beam and $a$ is a coefficient describing the pump induced losses. 
In our OPO$_{\rm{1}}$ we estimated $L_0$=0.00254 and $a$=0.00922 ($\mathrm{W}^{-1}$), respectively. These results yield $\rho = 0.97$. 
The noise level given by Eq.(\ref{Eq:SQ}) is modified as
\begin{equation}
{R^\prime}_\pm={R_\pm}{\cos^{2}\tilde{\theta}}+{R_\mp}{\sin^{2}\tilde{\theta}},
\label{Eq:SQPhase}
\end{equation}
due to the effect of phase fluctuations $\tilde{\theta}$ between the squeezed light and local oscillator in the balanced homodyne measurement, estimated to be $\tilde{\theta}$ =1.5 degree. Finally, 
\begin{equation}
{R^{\prime\prime}_\pm}={R^\prime_\pm}\left(1-10^{-\frac{C}{10}}\right) +10^{-\frac{C}{10}}.
\label{Eq:SQClear}
\end{equation}
where $C$ represents the clearance in dB of the shot noise level above the circuit noise. 
We estimated $C$ = 13.5 dB with a local oscillator power of 3.5 mW. 
By using the experimental parameters described above, we calculate the theoretically expected value of the noise level as displayed by the solid lines in Figure 2(b).
The calculation results show excellent agreement with the experimental values.

\end{document}